\pdfoutput=1
\documentclass[journal]{IEEEtran}

\renewcommand{\baselinestretch}{1}

\usepackage{tabularx,amsfonts,amssymb,amsmath}
\usepackage[dvips]{color}
\usepackage{mathrsfs}
\usepackage{subfigure}
\usepackage{calc,ifthen}
\usepackage{multirow}
\usepackage{bm}
\usepackage{float}
\usepackage{pstricks}
\usepackage{amsthm}
\usepackage{color}
\usepackage{empheq}
\usepackage{setspace}

\usepackage{cite}
\usepackage{psfrag}
\usepackage{pstool}
\usepackage{changes}
\usepackage{mathtools}
\usepackage{cuted}
\usepackage{gensymb}
\usepackage{color, colortbl}

\newcommand{\mynote}[1]{}
\newcommand{\newnote}[1]{}
\newcommand{\cancelled}[1]

\title{\LARGE \bf
Feedforward PID Control of Full-Car with Parallel Active Link Suspension for Improved Chassis Attitude Stabilization
}
\author{Zilin Feng, Min Yu, Simos A. Evangelou, Imad M Jaimoukha and Daniele Dini
\thanks{Zilin Feng ({\tt\small
zilin.feng17@imperial.ac.uk}), S. A. Evangelou {\tt\small (s.evangelou@imperial.ac.uk)} and Imad M Jaimoukha{\tt\small (i.jaimouka@imperial.ac.uk)} are with the Dept. of Electrical and Electronic at Imperial College London, UK .}
\thanks{Min Yu {\tt\small
(m.yu14@imperial.ac.uk)} and Daniele Dini {\tt\small
(d.dini@imperial.ac.uk)} are with the Dept. of Mechanical Engineering at Imperial College London, UK.}}
\begin{document}
\bstctlcite{IEEEexample:BSTcontrol}

\maketitle

\begin{abstract}

PID control is commonly utilized in an active suspension system to achieve desirable chassis attitude, where, due to delays, feedback information has much difficulty regulating the roll and pitch behavior, and stabilizing the chassis attitude, which may result in roll over when the vehicle steers at a large longitudinal velocity. To address the problem of the feedback delays in chassis attitude stabilization, in this paper, a feedforward control strategy is proposed to combine with a previously developed PID control scheme in the recently introduced Parallel Active Link Suspension (PALS). Numerical simulations with a nonlinear multi-body vehicle model are performed, where a set of ISO driving maneuvers are tested. Results demonstrate the feedforward-based control scheme has improved suspension performance as compared to the conventional PID control, with faster speed of convergence in brake in a turn and step steer maneuvers, and surviving the fishhook maneuver (although displaying two-wheel lift-off) with 50\,mph maneuver entrance speed at which conventional PID control rolls over.

\end{abstract}

\section{INTRODUCTION}
The suspension system refers to the entire support system consisting of springs and shock absorbers between the vehicle body and tires. The function of the suspension system is to support the body, and improve ride comfort and road holding\,\cite{jazar2017vehicle}. The passive suspension is a combination of mechanical components such as springs and dampers, which can only store or dissipate energy. In the modern car market with fierce competition, the performance of passive suspension is beginning to have difficulty to meet customers' demand in more efficient systems and growing requirements in high-quality ride comfort and road holding automobile performance. Active suspensions started to appear with chassis leveling, vibration attenuation, as well as low energy costs and high reliability requirements, which are compatible with the future electric vehicles.

Many advanced control approaches have been proposed for automobile semi-active and active suspension systems in the past few decades and achieved substantial results. The optimized sliding mode control algorithm combining linear quadratic optimal control algorithm is proposed in\,\cite{chen2017improved} to improve active suspension performance. In\,\cite{zhang2020active}, an active suspension control method based on the multi-agent prediction algorithm is proposed to attenuate the vertical acceleration of the suspension body. A robust optimal control method is designed for the suspension system in\,\cite{bai2021robust} to improve the vehicle performances in terms of handling stability and riding comfort. An active suspension controller based on feedback control for half-vehicle model is proposed in \cite{yang2007feedforward} to improve the vehicle ride comfort. The Linear parameter varying (LPV) feedforward filter is designed with a preexisting full-vehicle LPV controller in \cite{zheng2021comfort} to improve the stability of a vehicle subject to driver-induced roll disturbances. The feedforward compensation control based on steering angle is proposed in \cite{fleps2018lpv} to apply an anti-roll moment in advance which prevents the rollover. Apart from the control methods, many mechanical structures of active suspension are widely studied in automotive industry. Bose was first to develop electromagnetic actuators to replace the conventional spring-dampers~\cite{van2013robust}. Audi A8 introduced the adaptive air suspension, which combines a high-end air suspension with controlled damping. It enables the luxury sedan to remain composed and smooth on all types of uneven road surfaces~\cite{adcock2017audi}. The adaptive M suspension developed by BMW use electronically controlled shock absorbers and sensors to calculate the optimal damping forces in a few milliseconds~\cite{lischka2017business}. Mercedes-Benz introduced its Active Body Control (ABC) using hydro-pneumatic suspension~\cite{becker1996development}.

\begin{figure*}[htb!]
\centering
\begin{subfigure}[]
{
\centering
\includegraphics[width=0.8\columnwidth]{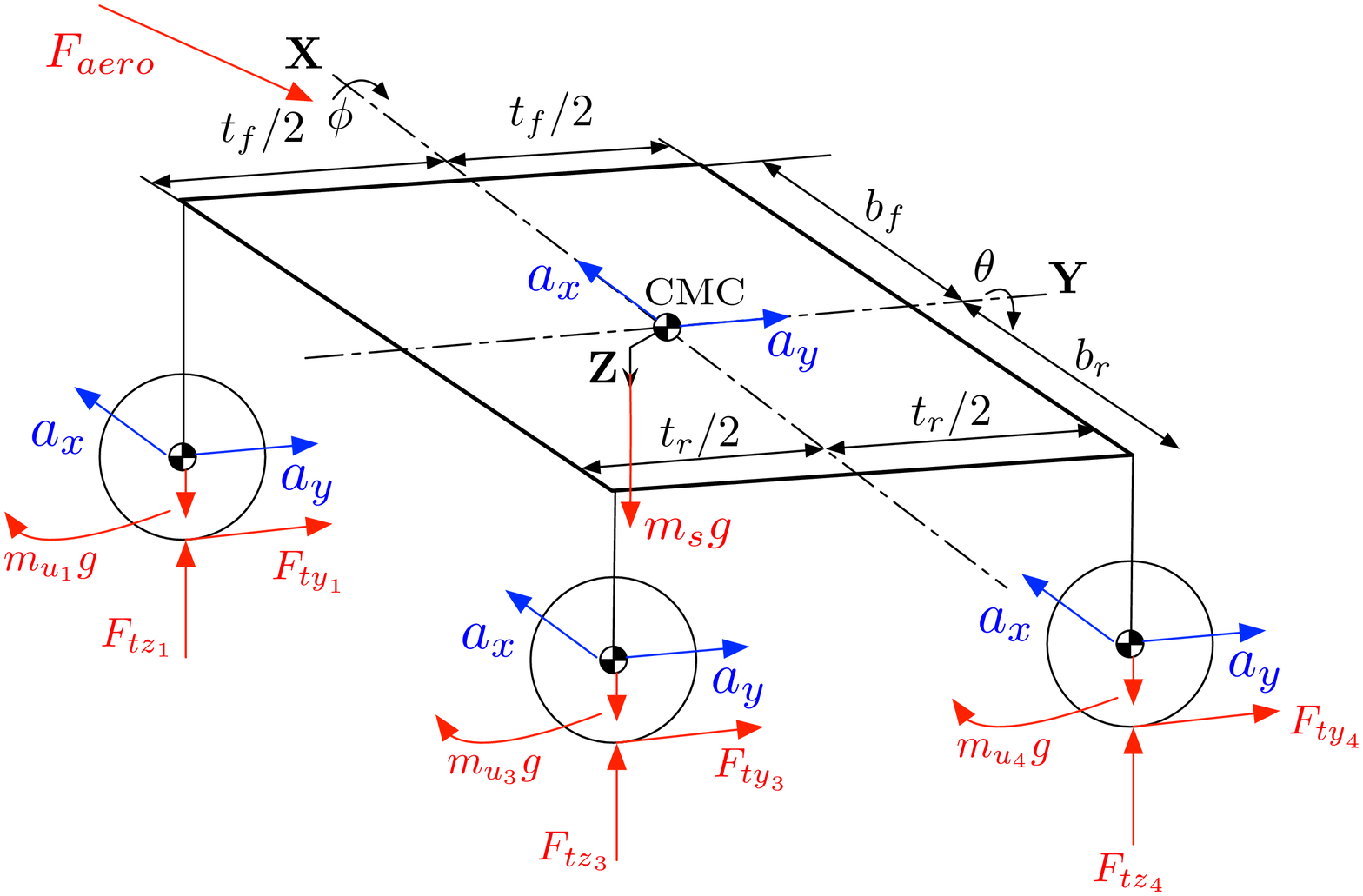}
}
\end{subfigure}\hspace{0.03\textwidth}
\begin{subfigure}[]{
\centering
\includegraphics[width=0.5\columnwidth]{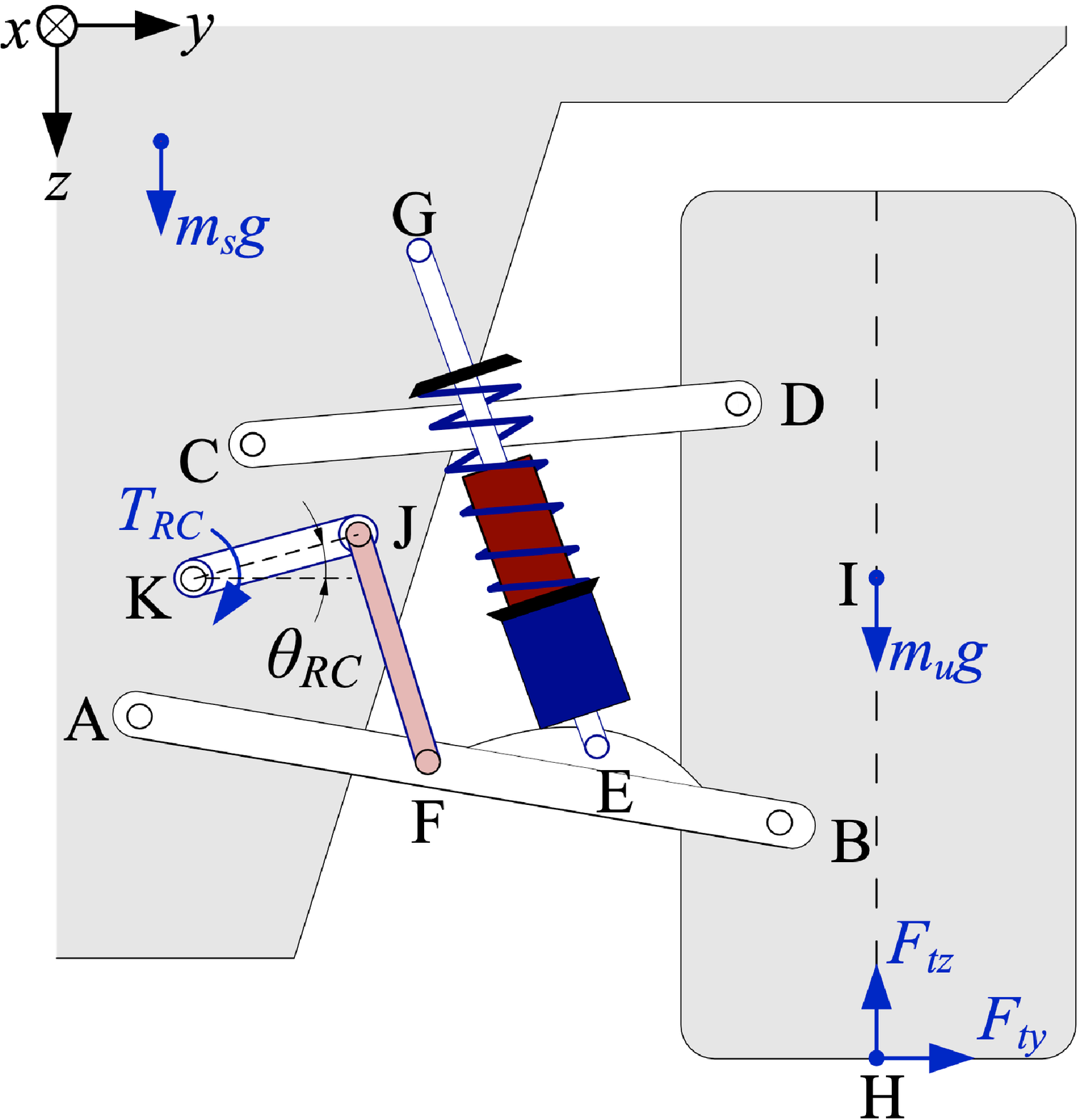}}
\end{subfigure}
\vspace{-2mm}
\caption{(a) Schematic of steady state PALS-retrofitted full car model. CMC corresponds to the center of the mass of the chassis, $m_s$ to the sprung mass, $m_{ui}$ to the unsprung mass at each corner, $b_f$ and $b_r$ to the front and rear wheel base, $t_f$ to the track width, $F_{aero}$ to the aerodynamic force. $F_{{ty}_i}$ and $F_{{tz}_i}$ are the lateral and vertical tire forces at each corner, respectively.(b) PALS application to the front-right corner of full car double-wishbone suspension. $T_{RC}$ is the rocker torque and $\theta_{RC}$ is the rocker angle\,\cite{yu2018parallel,9646263}.}
\label{fig1-1}
\end{figure*}

Recently, a novel mechatronic suspension solution, the Parallel Active Link Suspension (PALS), has been proposed in~\cite{yu2018parallel,9646263}. As shown in Fig.\,\ref{fig1-1}, in the PALS the rocker-pushrod assembly (`K-J-F') is introduced between the chassis and the lower wishbone ('A-B'), and in parallel with the spring-damper unit ('G-E'). The active segment the rocker (`K-J'), is driven by a rotary permanent magnet synchronous motor (PMSM) actuator, which delivers a torque ($T_{RC}$) acting from the chassis onto the lower wishbone (via the pushrod) to improve the performance of a double-wishbone suspension. The PALS features i) negligible unsprung mass and small sprung mass increment, ii) low power actuation requirements, with the efficient influence on the vertical tire force increment due to PALS links geometrically optimization and mechanical implementation, iii) simplified structure by the replacement of the anti-roll bar, iv) fail-safe characteristics and v) employment of developed and compact rotary-electromehcanical-actuation mechanics.\\\vspace{-2mm}

\vspace{-2mm}

Previous work develops the steady-state analysis for the PALS potential in terms of the chassis leveling and proposes the multi-objective PID control for the 
minimization of both the roll and pitch angles \cite{9646263}. However, despite these achievements, the previous work does not account for the delay due to the feedback control, which may result in the deterioration of the chassis attitude stabilization. This paper builds two compensation models to compensate the vertical tire force and tackle the feedback control delay through applying the feedforward control with an accurate nonlinear simulation fit based compensation model utilized. The main contributions of this paper are: i)\,development of two compensation models based on full car steady state analysis and nonlinear simulation fitting, respectively, and selection of the best performing compensation model by comparison of their fitting accuracy, ii)\,the design of `feedforward-PID' control that combines the nonlinear simulation fit based compensation model developed above with regular PID control, iii)\,numerical simulations with a nonlinear multi-body model of the PALS-retrofitted SUV full car to assess the effectiveness and robustness of the proposed control scheme, as compared to the passive suspension and the actively controlled PALS by conventional robust control, while the vehicle undergoes different ISO-defined road events. The rest of the paper is organized as follows: Section II illustrates the derivation of the feedforward normal load compensation with polynomial fitting method applied in the nonlinear simulation environment, and with steady state analysis of the full car. Section III introduces a feedforward PID control scheme based on nonlinear simulation fit based compensation model. Section IV performs numerical simulations to assess the performance of the proposed control scheme by comparing the proposed feedforward-PID scheme with conventional PID control in terms of pitch and roll angle minimization. Finally, concluding remarks are discussed in Section V.

\section{Development of tire force compensation model}\label{sec:model}
In this section, the mathematical steady-state model of the PALS-retrofitted full car that is developed in\,\cite{9646263} is briefly illustrated first, which motivates the proposal of one of the feedforward compensation laws. The other feedforward compensation law is introduced based on the relationship obtained by applying polynomial fitting through simulation of nonlinear full car model.

\subsection{Full car Steady state model and analysis}
As it can be shown in Fig.\,\ref{fig1-1}, the vertical tire force $F_{tz}$ at any given corner can be estimated as follows:
\begin{equation}\label{1-1}
\begin{aligned}
F_{tz} = F_{tz}^{(ne)} + \Delta F_{tz}^{(ax)} + \Delta F_{tz}^{(ay)} + \Delta F_{tz}^{(ae)},
\end{aligned}
\end{equation}
where $F_{tz}^{(ne)}$ is the vertical tire force in the nominal configuration, $\Delta F_{tz}^{(ax)}$ and $\Delta F_{tz}^{(ay)}$ represent the vertical tire force increments due to longitudinal and lateral acceleration, respectively, and $\Delta F_{tz}^{(ae)}$ is the force increment caused by the change in aerodynamic force $F_{tz}^{(aero)}$.

For the equilibrium of vertical forces to hold for the whole vehicle and for each axle independently, the following relationships must be satisfied:
\begin{equation}\label{1-2}
\begin{aligned}
&\Delta F_{{tz}_1}^{(ax)} = \Delta F_{{tz}_2}^{(ax)} = -\Delta F_{{tz}_3}^{(ax)} = - \Delta F_{{tz}_4}^{(ax)},\\
&\Delta F_{{tz}_1}^{(ay)} = -\Delta F_{{tz}_2}^{(ay)},\,\,\,\,
\Delta F_{{tz}_3}^{(ay)} = -\Delta F_{{tz}_4}^{(ay)},
\end{aligned}
\end{equation}
where the right subscript refers to the corner number 1 to 4 denoting the front left, front right, rear left and rear right corners, respectively. The vertical tire force increment due to vehicle longitudinal acceleration is obtained by the balance of pitching moments $\Sigma M_y = 0$:
\begin{equation}\label{1-3}
\begin{aligned}
\Delta F_{{tz}_1}^{(ax)}\!=\![\frac{m_s h_{CMC}+2(m_{{u}_f}R_{{wh}_f}+m_{{u}_r}R_{{wh}_r})}{2(b_f+b_r)}] a_x,
\end{aligned}
\end{equation}
where $m_{{u}_f}$ and $m_{{u}_r}$ correspond to the front and rear unsprung mass, respectively, which are not supported by the springs, $m_{s}$ to the sprung mass which is the rest of the mass of the car, $h_{CMC}$ to the height above the ground of the center of the mass of the chassis, $b_f$ and $b_r$ to the front and rear wheelbase and $R_{{wh}_f}$ and $R_{{wh}_r}$ to the front and rear wheel radius, respectively.

Similarly, the vertical tire force increment influenced by lateral acceleration can be estimated through the balance of rolling moments:
\begin{equation}\label{1-4}
\begin{aligned}
&M_x=[(m_sh_{CMC}+2(m_{{u}_f}R_{{wh}_f}+m_{{u}_r}R_{{wh}_r})]a_y,\\ 
&\Delta F_{{tz}_1}^{(ay)}t_f=(1-\sigma)M_x,\\
&\Delta F_{{tz}_3}^{(ay)}t_r=\sigma M_x,
\end{aligned}
\end{equation}
where $\sigma \in \left [0\,\,1 \right ] $ is defined as the ratio of overturning distribution (OCD) provided by the rear axle.

The vertical tire force increment on both front and rear axles caused by their aerodynamics force can be calculated through:
\begin{equation}\label{1-5}
\begin{aligned}
\Delta F_{{tz}_1}^{(ae)}\!=\!F_{{tz}_2}^{(ae)}\!=\!\frac{1}{2}c_{{ad}_f}{v_x^2},\\
\Delta F_{{tz}_3}^{(ae)}\!=\!F_{{tz}_4}^{(ae)}\!=\!\frac{1}{2}c_{{ad}_r}{v_x^2},\\
\end{aligned}
\end{equation}
where $c_{{ad}_f}$ and $c_{{ad}_r}$ are the aerodynamic downforce coefficients for the front and rear axles, respectively, and $v_x$ is the longitudinal velocity of the vehicle.

Based on the steady state vertical tire forces of the full car provided in (\ref{1-1})-(\ref{1-5}), the chassis leveling capability of PALS can be achieved 
as follows.

Applying the principle of virtual work to a corner of the car with rocker-pushrod and the double wishbone linkages with the road wheel taken into account and considering a static chassis yields
\begin{equation}\label{1-6}
\begin{aligned}
T_{{RC}_i} = \frac{\partial z_{H_i}}{\partial \theta_{{RC}_i}} \Delta F_{{tz}_i} = \frac{\partial l_{s_i}}{\partial \theta_{{RC}_i}} \Delta F_{{RC}_i},
\end{aligned}
\end{equation}
where $z_{H_i}$ is the vertical coordinate of the road wheel center, $\theta_{{RC}_i}$ is the rocker angle with respect to the horizontal line, $l_{s_i}$ is the suspension deflection, and $F_{{RC}_i}$ is the linear equivalent actuation force.

Transformation functions between $\Delta F_{{tz}_i}$ and $T_{{RC}_i}$ at each corner can be defined as follows:
\begin{equation}\label{1-50}
\begin{aligned}
\beta_{i} = \beta_{i}(z_{H_i}) = \frac{\Delta F_{{tz}_i}}{T_{{RC}_i}}  = \frac{\partial \theta_{{RC}_i}}{\partial z_{H_i}}\cdot
\end{aligned}
\end{equation}



For the feedback control system, the controller does not work well until the deviation of the roll angle reaches an ultimate value. However, it is hard to regulate the system when the roll angle deviation is large or the roll rate is increasing which may result in rollover. Therefore, the vehicle's longitudinal and lateral acceleration shown in \eqref{1-3}-\eqref{1-4} are estimated according to the steering wheel angle, longitudinal speed, gas pedal and brake pedal position. Then, the active suspension produces the vertical tire force in advance to ensure that both roll and pitch angle are minimized.

\subsection{Nonlinear simulation fit based compensation model}\label{sec:model2}
\begin{figure}[htb!]
\begin{center}
\includegraphics[width=1.0\columnwidth]{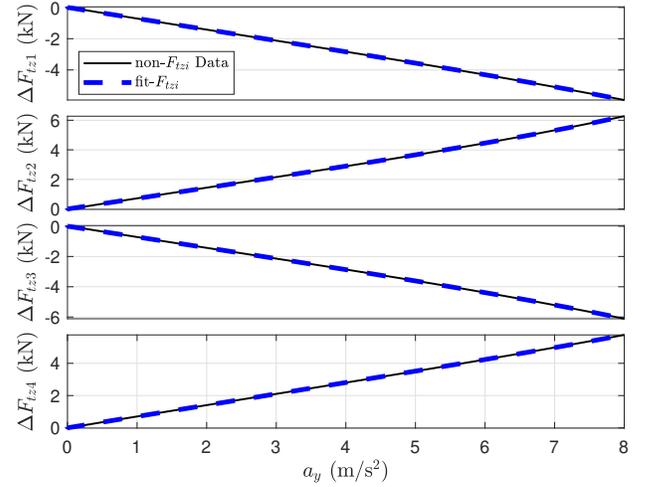}    
\caption{Vertical tire force increment ($\Delta F_{tzi}$) polynomial fitting with respect to the lateral acceleration ($a_y$) in the steady state cornering maneuver, where the black solid line corresponds to the nonlinear simulation $\Delta F_{tzi}$ data and the blue dashed line to the third order polynomial fitting line}
\label{fig1-4}
\end{center}
\end{figure}

\begin{figure}[htb!]
\begin{center}
\includegraphics[width=1.0\columnwidth]{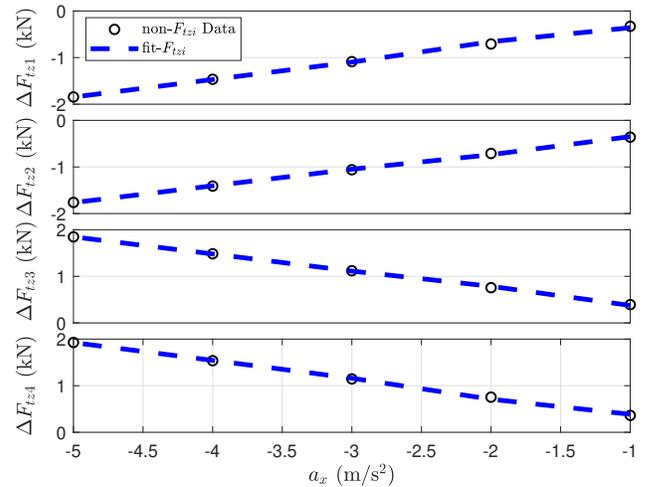}    
\caption{Vertical tire force increment ($\Delta F_{tzi}$) linear fitting with respect to the longitudinal acceleration ($a_x$) in the constant deceleration maneuver, where the black circle corresponds to the nonlinear simulation $\Delta F_{tzi}$ data and the blue dashed line to the first order polynomial fitting line}
\label{fig1-5}
\end{center}
\end{figure}

In addition to the steady state model utilized in\,\cite{9646263}, a new practical approach is introduced in this section which uses a polynomial fitting method in the nonlinear simulation of a high fidelity nonlinear multi-body model developed in\,\cite{9646263} to establish the relationship between longitudinal acceleration ($a_x$), lateral acceleration ($a_y$) and the vertical tire force increments ($\Delta F_{tzi}$) at each corner. The approach is detailed as follows: i) The steady state cornering maneuver is selected to find the relationship between $\Delta F_{tzi}$ and $a_y$ only, due to the linear growth of $a_y$ with constant $v_x$; ii) similarly, with initial longitudinal speed $v_x$\,=\,100\,km/h, the straight line constant deceleration maneuver is applied to establish the relationship between $\Delta F_{tzi}$ and $a_x$. The plots of the aforementioned $\Delta F_{tzi}$ with regards to $a_y$ and $a_x$ are shown in Fig.\,\ref{fig1-4} and Fig.\,\ref{fig1-5}, respectively.

To achieve the best fitting, $\Delta F_{tzi}^{(a_y)}$ is selected as a third order polynomial fitting and $\Delta F_{tzi}^{(a_x)}$ is chosen as first order polynomial fitting. Hence, the total vertical tire force increment of the selected compensation model is derived as follows:
\begin{equation}
\begin{aligned}
\Delta F_{tzi} = \Delta F_{tzi}^{(a_x)}+\Delta F_{tzi}^{(a_y)}.
\end{aligned}
\end{equation}
\vspace{-2mm}

The comparison of vertical tire force increments $\Delta F_{tzi}$ between the steady state model and the nonlinear simulation fit model in terms of steady state cornering and brake in turn ISO maneuvers are shown below in Fig.\,\ref{fig1-20} and Fig.\,\ref{fig1-21}. 
As it can be seen in these two plots, although the steady state model provides the information on $\Delta F_{tzi}$ through  \eqref{1-1}-\eqref{1-5}, the nonlinear simulation fit model offers more accurate results for $\Delta F_{tzi}$. Therefore, the nonlinear simulation fit model is selected as the $\Delta F_{tzi}$ compensation model. 

\vspace{-2mm}

\vspace{-2mm}

\section{Control Methodology Development}\label{sec:control}
This section describes the synthesis of the two control schemes with PALS for the chassis leveling contributed by the present work: A)\,the previously proposed multi-objective PID control \cite{9646263}, and B)\,the newly proposed feedforward PID control based on nonlinear simulation fit, which is detailed in sec.\ref{sec:model2}.

\begin{figure}[htb!]
\begin{center}
\includegraphics[width=1.0\columnwidth]{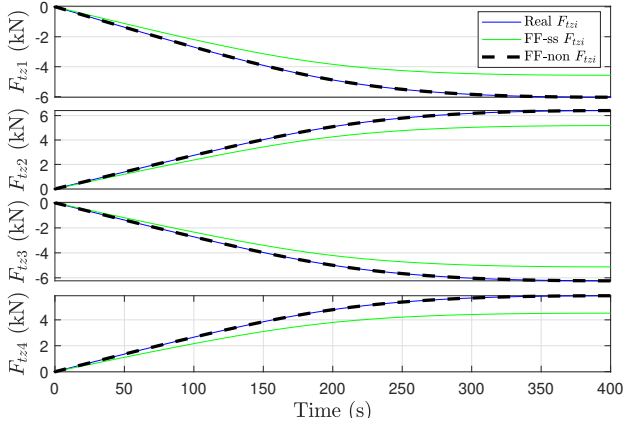}    
\vspace{-2mm}
\caption{Vertical tire force increment ($\Delta F_{tzi}$) with steady state model and nonlinear simulation fit model when the vehicle is driven over an ISO steady-state cornering at 100\,km/h ($a_y$ from 0\,m/s$^2$ to 8\,m/s$^2$)}
\vspace{-2mm}
\label{fig1-20}
\end{center}
\end{figure}
\vspace{-2mm}
\begin{figure}[htb!]
\begin{center}
\includegraphics[width=1.0\columnwidth]{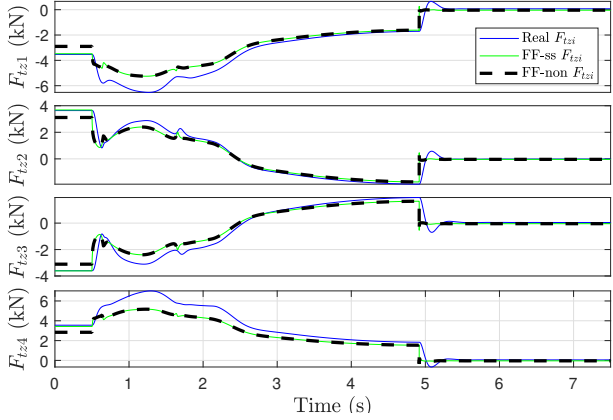}    
\vspace{-2mm}
\caption{Vertical tire force increment ($\Delta F_{tzi}$) with steady state model and nonlinear simulation fit model when the vehicle is driven over an ISO brakeinturn at initial longitudinal speed 80\,km/h}
\vspace{-2mm}
\label{fig1-21}
\end{center}
\end{figure}
\vspace{-2mm}

\subsection{PID control (`PALS-PID')}\label{sec:control1}
\begin{figure}[htb!]
\begin{center}
\includegraphics[width=1.0\columnwidth]{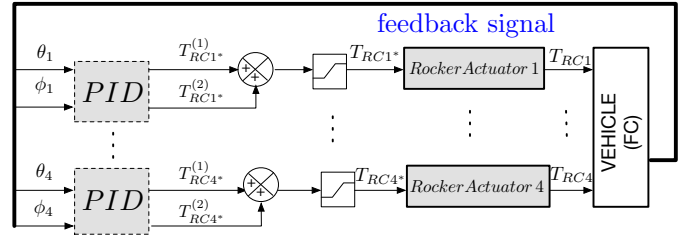}    
\vspace{-2mm}
\caption{Multi-objective PID control implementation in the PALS-retrofitted full car.\,\cite{Aranaphdthesis,9646263}}
\label{fig1-6}
\end{center}
\end{figure}

To achieve desirable chassis attitude and driving dynamics, the multi-objective PID control scheme (`PALS-PID') is adapted to the PALS-retrofitted full car\,\cite{9646263}. As it can be shown in Fig.\,\ref{fig1-6}, a group of PID controllers at each corner $i$, are synthesized to deal with pitch and roll angle control, with $\theta_{i}$ and $\phi_{i}$ denoting feedback signals. The reference rocker torque $T_{{RCi}^*}$ with pitch angle minimization, at each corner is obtained as follows:
\vspace{-2mm}
\begin{equation}\label{1-17}
\begin{aligned}
T_{{RC1}^*}^{(1)} &= T_{{RC2}^*}^{(1)} = -K_{p,f}^{(1)}{\theta} - K_{i,f}^{(1)}\int{\theta} - K_{d,f}^{(1)}\dot{\theta},\\
T_{{RC3}^*}^{(1)} &= T_{{RC4}^*}^{(1)} = K_{p,r}^{(1)}{\theta} + K_{i,r}^{(1)}\int{\theta} + K_{d,r}^{(1)}\dot{\theta},
\end{aligned}
\end{equation}
and the reference rocker torque with roll angle minimization is:
\begin{equation}\label{1-18}
\begin{aligned}
T_{{RC1}^*}^{(2)} &= -T_{{RC2}^*}^{(2)} = -K_{p,f}^{(2)}{\phi} - K_{i,f}^{(1)}\int{\phi} - K_{d,f}^{(2)}\dot{\phi},\\
T_{{RC3}^*}^{(2)} &= -T_{{RC4}^*}^{(2)} = -K_{p,r}^{(2)}\dot{\phi} - K_{i,r}^{(2)}\int{\phi} - K_{d,r}^{(2)}\dot{\phi},
\end{aligned}
\end{equation}
with the PID tuning parameters detailed in Table \ref{tab1-2}. The overall reference rocker torque feeding the rotary rocker actuator at each corner is:
\begin{equation}\label{1-19}
\begin{aligned}
T_{{RCi}^*} = T_{{RCi}^*}^{(1)} + T_{{RCi}^*}^{(2)}.
\end{aligned}
\end{equation}

Through the inner-loop tracking control at each corner, the $T_{{RCi}^*}$ is transformed to $T_{RCi}$ and then fed to the vehicle system which is detailed in\,\cite{9646263}.

\vspace{-2mm}

\subsection{Nonlinear Feedforward PID (`FF-PID-non')}\label{sec:control2}
\begin{figure}[htb!]
\begin{center}
\includegraphics[width=1.0\columnwidth]{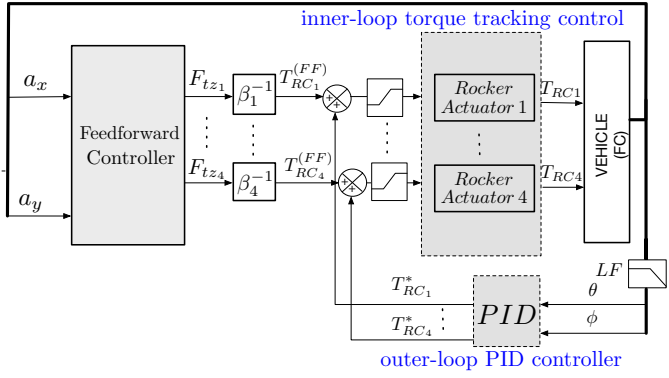}    
\vspace{-2mm}
\caption{`FF-PID-non' control implementation in the PALS-retrofitted full car with PID block defined in Fig.\,\ref{1-6}.} 
\label{fig1-8}
\end{center}
\end{figure}

The previous multi-objective PID control loop regulates the output pitch and roll angle of the vehicle plant employing negative feedback. On the other hand, as shown in Fig.\,\ref{fig1-8}, with both longitudinal and lateral acceleration measured from the vehicle, the nonlinear Feedforward PID control approach utilizes the feedforward compensation law, proposed in Section~\ref{sec:model2}.A, to calculate the vertical tire force $F_{{tz}_i}$. Through the conversion block $\beta_{i}$ in \eqref{1-50}, the $T_{{RC}_i}^{(FF)}$ are obtained to achieve major compensation of the actuation torques of the rotary actuators. Usually, the feedforward control cannot directly compensate the full information of the output rocker torque $T_{RCi}$ and therefore requires combining with the PID feedback control loops. Then the inner loop rocker torque tracking control links the feedforward control and the mechanical system of the PALS full car, where d-q transformation and zero d-axis current controls proposed in\,\cite{9646263} are utilized, such that the three-phase PMSMs behave as DC equivalent motors, with the produced torques proportional solely to the q-axis currents. 

It is notable that the PID parameters should be retuned in nonlinear feedforward PID (listed in Table~\ref{tab1-2} in the Appendix) as compared to those in `PALS-PID' due to the effect of the feedforward control. However, this tuning  for PID is time-efficient since feedforward control accounts for nearly 90\% of the total $T_{RCi}$, making the PID contribute only marginally to the $T_{RCi}$  regardless of aggressive tuning selection.

\vspace{-2mm}

\section{Numerical Simulations with nonlinear multi-body model and analysis}
In this section, with the nonlinear multi-body model described in\,\cite{9646263} and the control strategies proposed in Section\,\ref{sec:control}, a group of ISO driving maneuvers, containing, A)\,step steer, B)\,steady-state cornering, C)\,brake in a turn, D)\,pure longitudinal braking and acceleration, E)\,fishhook, and F)\,continuous sinusoid steer, are tested to evaluate the efficiency and robustness of the synthesized controllers ('PALS-PID' synthesized in Section\,\ref{sec:control1} and 'FF-PID-non' synthesized in Section\,\ref{sec:control2}).

\subsection{Step Steer}
\begin{figure}[htb!]
\begin{center}
\includegraphics[width=8cm]{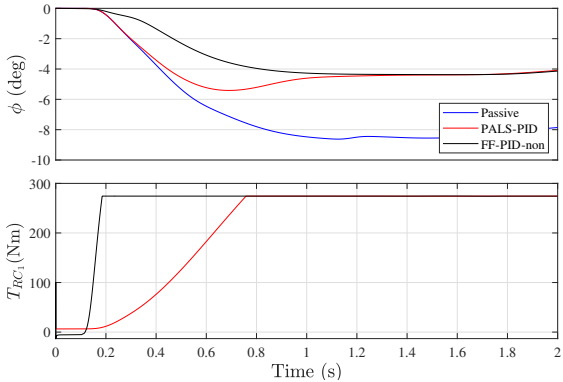}    
\caption{Numerical simulation results: the chassis roll angle ($\phi$) and front left corner rocker torque ($T_{{RC}_1}$) when the vehicle is driven over an ISO step steer at 100\,km/h, for different methods of suspension control.}
\label{fig1-9}
\end{center}
\end{figure}

ISO 7401:2011 details an open-loop test method to determine the transient response behavior of passenger vehicles\,\cite{iso2011road}. Here, the PALS-retrofitted full car is driven at a constant forward speed of 100\,km/h, with the steering wheel angle increasing at a constant rate of 500\,deg/s from\,\mbox{0\,$\degree$} to \mbox{48.6\,$\degree$} such that the vehicle stabilizes at a lateral acceleration of $a_y$\,=\,8\,m/s$^2$. The reduction in roll angle $\phi$ achieved thanks to the synthesized controllers is presented in Fig.\,\ref{fig1-9}, where `PALS-PID' provides 42\% mitigation of roll angle RMS value over the passive suspension. The `FF-PID-non' produces even better performance in terms of roll angle RMS value attenuation than `PALS-PID', with 52\% mitigation of roll angle over the passive suspension. Furthermore, it does not suffer as the `PALS-PID' does from a positive roll angle slope at approximately 0.7\,s, and as Fig.\,\ref{fig1-9} shows, the response of `FF-PID-non' is much faster than that of `PALS-PID' in terms of front-left corner rocker torque $T_{{RC}_1}$.

\vspace{-2mm}

\subsection{Steady State Cornering}
ISO 4138:2004 defines an open-loop test method to assess the potential of the passenger vehicles for roll mitigation in steady-state circular driving\,\cite{iso20124138}.

\begin{figure}[htb!]
\begin{center}
\includegraphics[width=8cm]{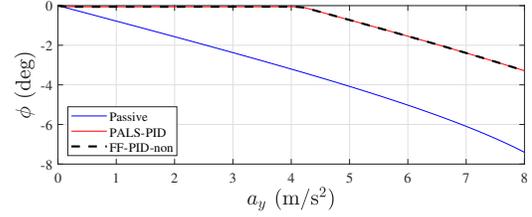}    
\caption{Numerical simulation results: the chassis roll angle ($\phi$) when the vehicle is driven over an ISO steady-state cornering at 100\,km/h, for different cases of active suspension control.}
\label{fig1-10}
\end{center}
\end{figure}

The vehicle is driven at a constant longitudinal speed of 100\,km/h, with the angle of the steering wheel linearly increased from \mbox{0\,$\degree$} to \mbox{60\,$\degree$} in 400\,s. Fig.\,\ref{fig1-10} depicts that the roll angle is completely neutralised up to lateral accelerations of approximately 4\,m/s$^2$ for `PALS-PID'. The `FF-PID-non' control strategy presents the same performance enhancement over the passive suspension as `PALS-PID', due to the successful tracking of the saturating reference actuation forces.

\vspace{-2mm}

\subsection{Brake in a turn}
ISO 7975:2006 presents an open-loop test method for determining the steady-state circular response of a vehicle that is altered by a sudden brake. As defined in\,\cite{ISO_7975-2006}, the PALS-retrofitted full car is initially driven in a circular path of 100\,m radius at a constant lateral acceleration of 5\,m/s$^2$,  corresponding to a constant forward speed of 80\,km/h, then the steering wheel is fixed and brakes applied to enable the vehicle to slow down at a constant deceleration of $a_x\!=\!\mbox{-}$5\,m/s$^2$.

\begin{figure}[htb!]
\begin{center}
\includegraphics[width=8.4cm]{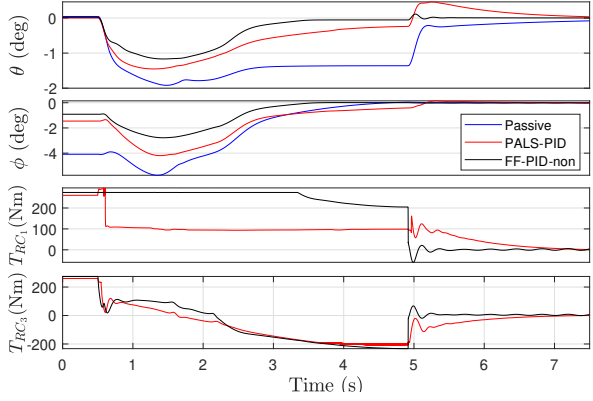}    
\caption{Numerical simulation results: the chassis pitch angle ($\theta$), chassis roll angle ($\phi$), rocker torque of front left corner  ($T_{{RC}_1}$) and rear left corner ($T_{{RC}_3}$) when the vehicle is driven over an ISO brake in turn at initial longitudinal speed 80\,km/h, for different methods of suspension control.}
\label{fig1-11}
\end{center}
\end{figure}

The variations of the roll angle, $\phi$, and pitch angle, $\theta$, are shown in Fig.\,\ref{fig1-11}. It can be seen that, as compared to the passive suspension, the `PALS-PID' has less overshoot but it takes a longer time to settle in terms of roll angle performance, and its average pitch angle is reduced from \mbox{-1.5\,$\degree$} to \mbox{-0.4\,$\degree$} with a larger overshoot at approximately 5\,s before it comes to a stop. `FF-PID-non' has a similar performance to that of `PALS-PID' with slightly smaller overshoot over an initial time period of 0-2\,s in terms of pitch angle performance, then it converges much faster to zero without suffering as the `PALS-PID' does from a positive pitch angle spike at approximately 5\,s when the car comes to a stop and restores its equilibrium position; and its roll angle is reduced with a much smaller overshoot at approximately 1.5\,s and shorter time to settle as compared to `PALS-PID'. Fig.\,\ref{fig1-11} also shows the plots of the rocker torque of front left corner  ($T_{{RC}_1}$) and rear left corner ($T_{{RC}_3}$). As it can be seen, the response of `FF-PID-non' is much faster than that of `PALS-PID' in terms of stopping the vehicle.

\vspace{-2mm}

\subsection{Pure longitudinal acceleration and braking}
\begin{figure}[htb!]
\begin{center}
\includegraphics[width=8cm]{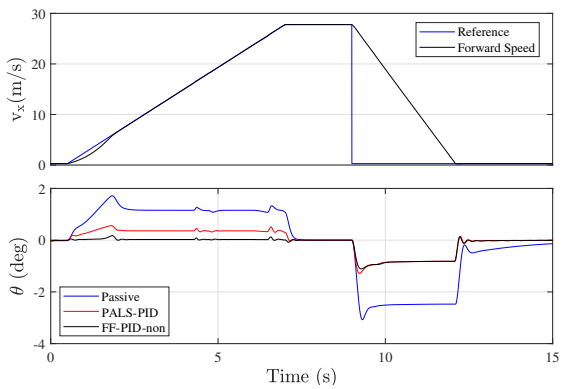}    
\caption{Numerical simulation results: the forward speed ($v_x$) and the chassis pitch angle ($\theta$) when the vehicle is driven over an ISO longitudinal acceleration/deceleration maneuver, for different methods of suspension control.}
\label{fig1-12}
\end{center}
\end{figure}

The pure longitudinal accelerating and braking maneuver is used to assess the ability of the controlled system for pitch mitigation. In this maneuver, a hard acceleration process starts from 1\,km/h to 100\,km/h in 6.5\,s, which is then followed by a 2\,s constant speed period and an emergency stop. Time responses for the PALS full car in this maneuver are shown in Fig.\,\ref{fig1-12}. In the top row, the reference and actual speed profiles are compared. As it can be seen, the car follows the acceleration profile closely, and the resulting deceleration rates remain within the 1.05\,g to 1.25\,g band during the emergency stop. Pitch angle simulation results for different controllers are displayed in the bottom row of Fig.\,\ref{fig1-12}, which shows that the `PALS-PID' is capable of maintaining an overall flat pitch angle during the acceleration phase and achieves approximately 35\%  pitch angle correction during the emergency stop. The `FF-PID-non' achieves almost $\frac{1}{3}$ of the average pitch angle during the acceleration phase and marginally better response during emergency stop as compared to `PALS-PID'. 

\vspace{-2mm}

\subsection{Fishhook}
The fishhook defines an open-loop test procedure to evaluate the vehicle dynamic rollover propensity\,\cite{national2013laboratory}.
\begin{figure*}
\centering
\begin{subfigure}
  \centering
  \includegraphics[width=.47\linewidth]{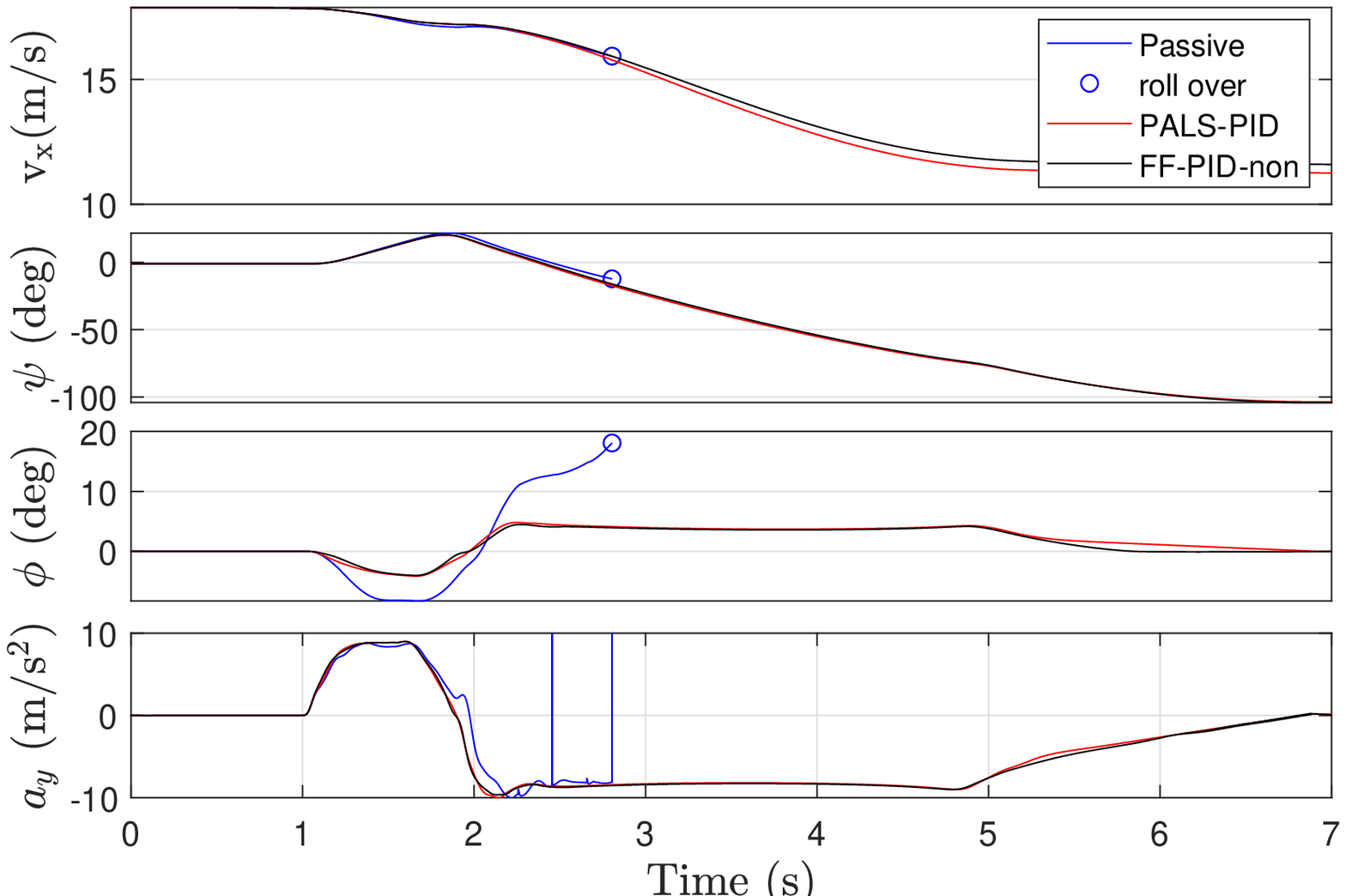}  
\end{subfigure}
\begin{subfigure}
  \centering
  \includegraphics[width=.47\linewidth]{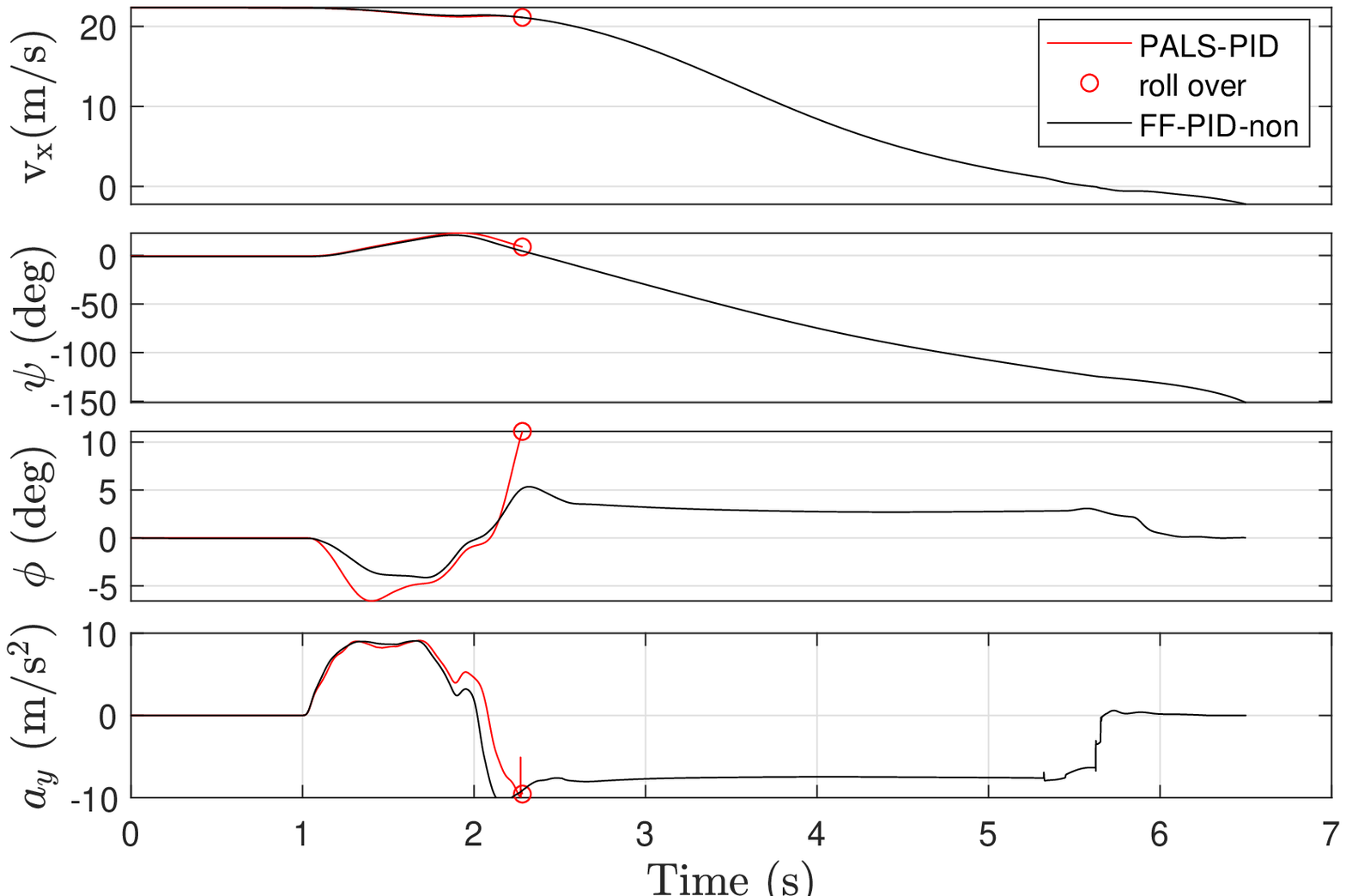}  
\end{subfigure}
\begin{subfigure}
  \centering
  \includegraphics[width=.47\linewidth]{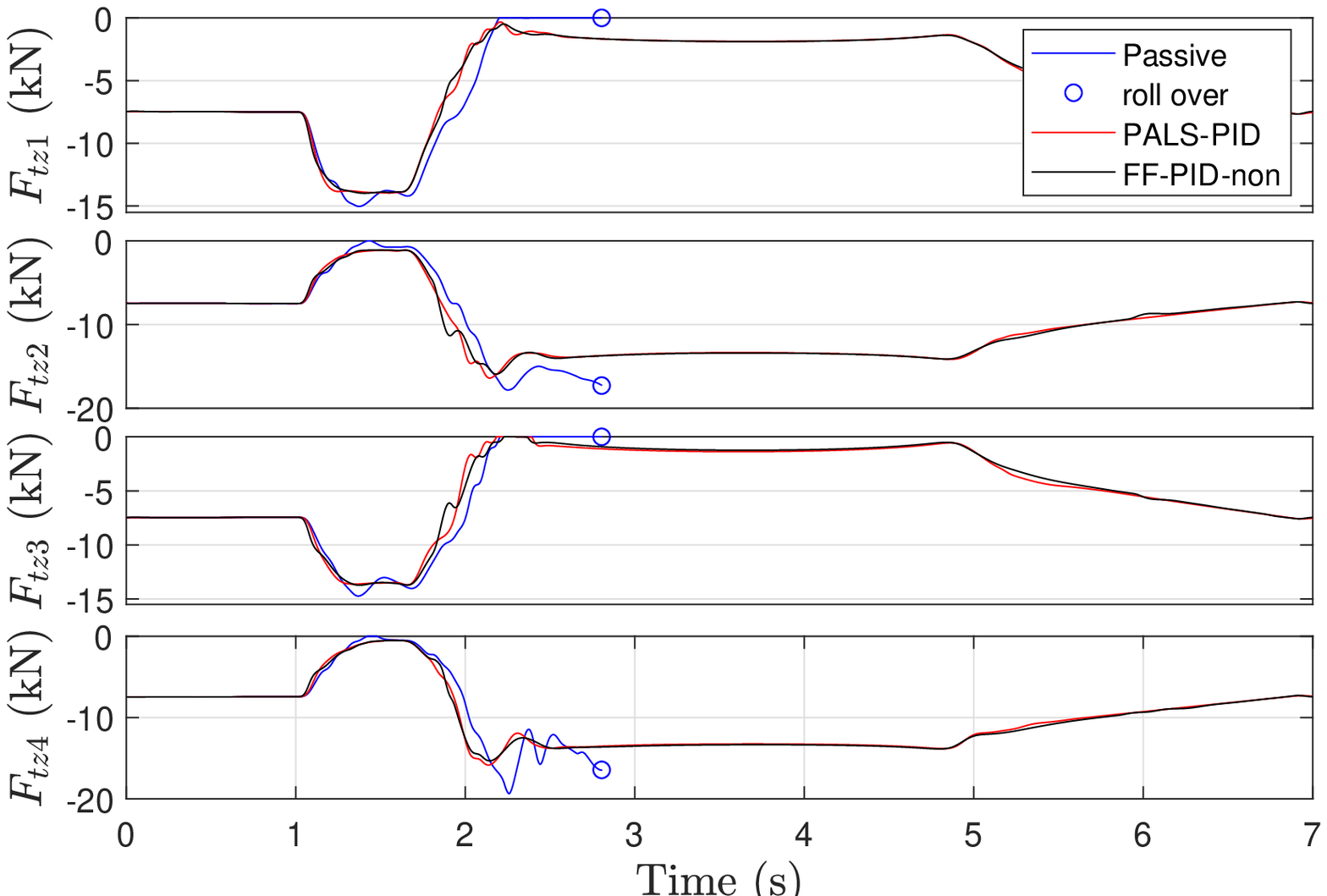}  
\end{subfigure}
\begin{subfigure}
  \centering
  \includegraphics[width=.47\linewidth]{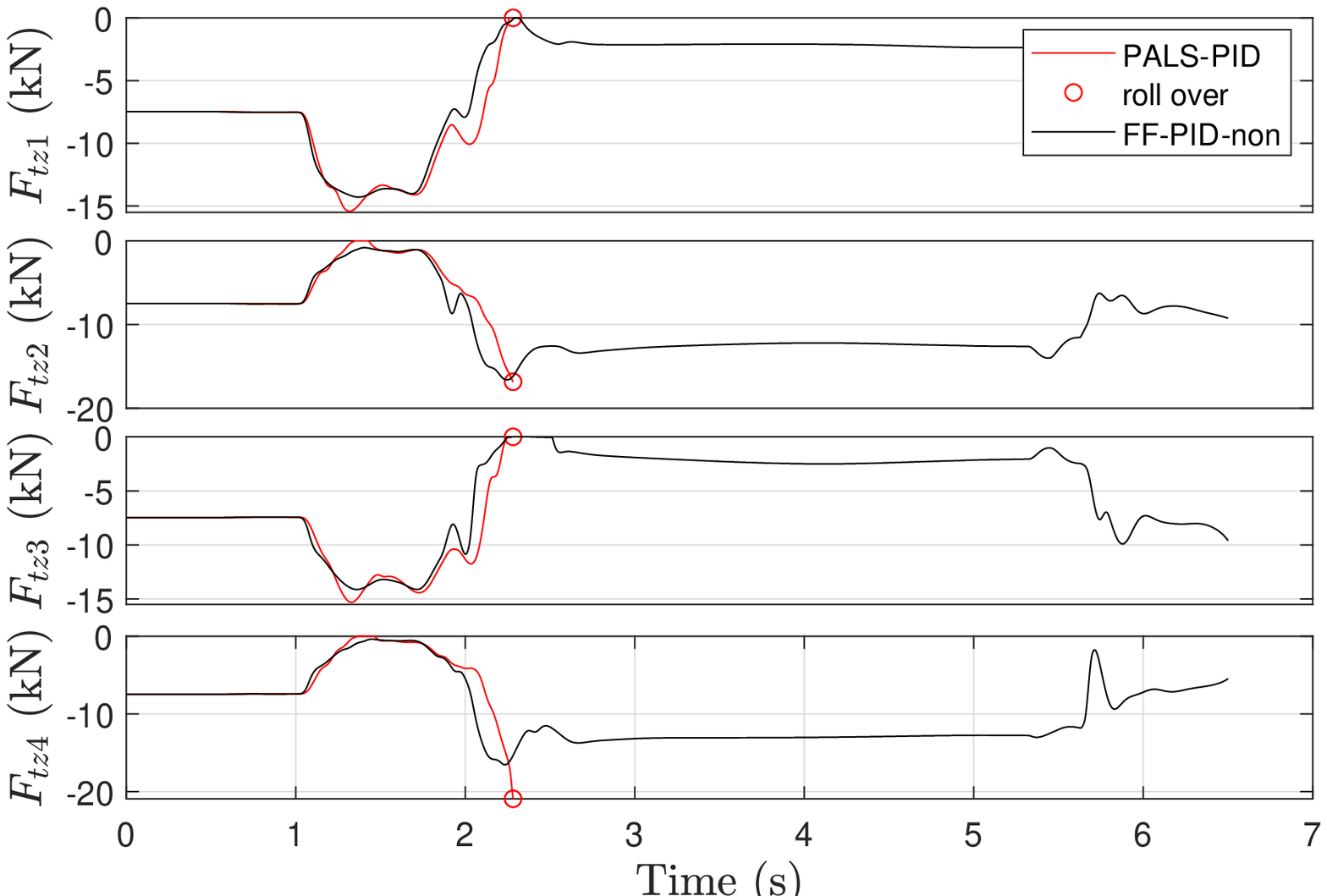}  
\end{subfigure}
\vspace{-2mm}
\caption{Numerical simulation results: 
the forward speed ($v_x$), chassis yaw angle ($\psi$), roll angle ($\phi$) and lateral acceleration ($a_y$) with MES=40\,mph (top-left), and MES=50\,mph (top-right); and the vertical tire force ($F_{tzi}$) with with MES=40\,mph (bottom-left), and MES=50\,mph (bottom-right) when the vehicle is driven over fishhook maneuvers for different methods of suspension control.}
\label{fig1-13} 
\vspace{-6mm}
\end{figure*}

The test procedure consists of two stages. In the first one, the PALS-retrofitted full car is driven at a forward speed of 50 miles per hour\,(mph), while a slow turning maneuver is performed to determine the steering wheel angle $\delta_{ini}$ required to reach a lateral acceleration of 0.3\,g. In the second stage, the vehicle is driven in a straight line at a certain Maneuver Entrance Speed\,(MES). Then, the throttle pedal is released, and the steering-wheel angle is rotated up to $\delta_{fh}=6.5\,\delta_{ini}$. The steering wheel angle is reversed to -\,$\delta_{fh}$ when the roll rate reduces below 1.5\,deg/s, staying there for 3\,s, after which it reverts to zero over a time period of 2\,s. The first manoeuvre is performed with MES=35\,mph, then it is repeated with MES=40,\,45 and 50\,mph.

The passive car survives the fishhook with MES\,=\,35mph, but suffers from rollover at MES\,=\,40mph. The PALS-retrofitted full car with `PALS-PID' control manages to avoid rollover at MES\,=\,40mph and MES\,=\,45mph, while it cannot maintain the roll angle stability at MES\,=\,50mph. In contrast, the feedforward control strategy `PALS-PID-non' is capable of presenting robust and stable performance for all four test values of MES.

Results for MES\,=\,40mph are presented in the first column of Fig.\,\ref{fig1-13}  for the passive and active configurations. In the top-left plot of Fig.\,\ref{fig1-13}, both active controllers keep the roll angle below 5 deg in the first steering phase, and the roll angle peaks at around 5 deg in the second. The forward speed, yaw angle and lateral acceleration are also included in the plots to help understand the overall performance. The results of vertical tire force at each corner with different controllers are shown in the bottom-left plot of Fig.\,\ref{fig1-13}. The passive suspension displays two-wheel lift at the point indicated by blue circles, and eventually rolls over. The other two active controllers also display two-wheel lift at t\,=\,2.23\,s, but regain contact with the ground quickly at t\,=\,2.29\,s and remain stable throughout the maneuver.

Results for MES\,=\,50mph are shown in the second column of Fig.\,\ref{fig1-14} for active configurations only. The chassis parameters including forward speed, yaw angle, roll angle and lateral acceleration are displayed in the top-right plot of Fig.\,\ref{fig1-13}. It is clear that `PALS-PID-non' keeps the roll angle below 5 deg throughout the maneuver. Two-wheel lift occurs at time t\,=\,2.26s, but contact with the ground is reclaimed at time t\,=\,2.38s and the vehicle remains stable until the end of the maneuver which can be verified in the bottom-right plot of Fig.\,\ref{fig1-13}. However, `PALS-PID' displays two-wheel lift at t\,=\,2.28s and finally rolls over.

\subsection{Continuous sinusoid steer}
The continuous sinusoid steer defines an open-loop test procedure to understand the performance of the PALS at various frequencies. 

\begin{figure}[htb!]
\begin{center}
\includegraphics[width=8cm]{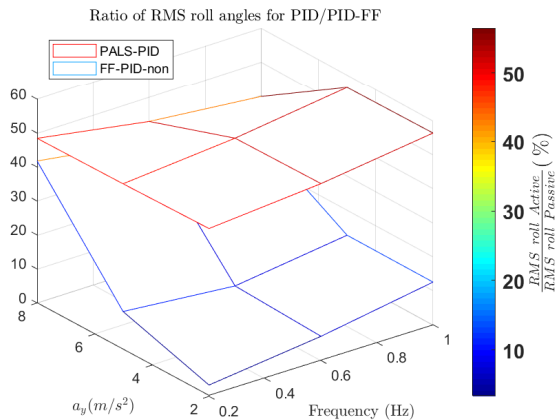}    
\caption{Numerical simulation results: Ratio of RMS roll angles obtained with different cases of active and passive suspensions when the vehicle is driven over an sinusoid steering-wheel maneuver.}
\label{fig1-14}
\end{center}
\end{figure}

A continuous steering-wheel sinusoid is applied when the vehicle is driven in a straight line at 100\,km/h as defined in\,\cite{iso2011road}. Steering frequencies from 0.2\,Hz to 1\,Hz in 0.2\,Hz steps are applied and results obtained with different cases of active suspensions are compared with the passive suspension. The ratios of the RMS roll angle obtained with the two active suspension methods over the one computed with the passive suspension are shown in Fig.\,\ref{fig1-14}. This ratio remains around 50\% in most of the cases for `PALS-PID', while the ratio for the feedforward controller remains below 20\% in most cases and rises to 40\% at the whole steering frequency range and large steering-wheel amplitudes which indicate better performance in terms of chassis leveling than `PALS-PID'.
\vspace{-2mm}

\section{Conclusion}
The recently proposed mechatronic suspension of the Parallel Active Link Suspension (PALS) is investigated in the application to a SUV full car\,\cite{9646263}, with feedforward compensation taken into consideration in the suspension control design, revealing promising potential for chassis leveling and stabilization.

The proposed feedforward PID control strategy with nonlinear polynomial fitting method applied is proposed for PALS low-frequency application, with essential improvement over the passive suspension system and decent enhancement as compared to `PALS-PID' in terms of chassis leveling and speed of convergence over a set of ISO driving maneuvers.

In future work, the PALS performance in higher-frequency road events is to be tested for the integration of $H_{\infty}$ control ($\mu$-synthesis control) developed elsewhere and the proposed feedforward PID control. The ride comfort and road holding related variables (CMC vertical acceleration and the tire deflection) require a more comprehensive assessment of the PALS performance before any on-road experiments.

\appendix
\begin{table}[htb!]
\centering
\caption{Main Parameters of Original and PALS Retrofitted SUV full car (F: Front, R: Rear)}\label{tab1-1}
\label{tab:vehicle data}
\scalebox{1}{\begin{tabular*}{1\columnwidth}{@{\extracolsep{\fill}}l @{\extracolsep{\fill}} c@{\extracolsep{\fill}}
c@{\extracolsep{\fill}}c @{\extracolsep{\fill}} c}
\hline
\hline
 Parameters & Value\\
 \hline
 \multicolumn{4}{c}{Original vehicle\,\cite{arana2016series,9646263}}\\
 \hline
 CMC Height & 0.71\,m \\
 F/R Weight distribution & 50\,/\,50\,$\%$ \\
 Total/Sprung mass & 2950\,/\,2700\,kg \\
 F/R OCD ratio\,(1\,-\,$\sigma$\,/\,$\sigma$) & 57\,/\,43\,$\%$\\
 F/R Spring stiffness & 53.5\,/\,53.1\,$\frac{kN}{m}$ \\
 F/R Tire stiffness & 290\,$\frac{kN}{m}$ \\
 F/R Tire damping & 300\,$\frac{Ns}{m}$ \\
 F/R Track width\,($t_f/t_r$) & 1.677\,/\,1.696\,$m$ \\
 F/R Wheelbase\,($b_f/b_r$) & 1.538\,/\,1.538\,$m$\\
 F/R Wheel radius\,($R_{{wh}_f}/R_{{wh}_r}$) & 0.385\,/\,0.385\,$m$\\
 F/R Unsprung mass\,($m_{uf}$\,/\,$m_{ur}$) & 62.5\,/\,62.5\,kg\\
 \hline
 \multicolumn{4}{c}{PALS retrofit (actuator per corner)\,\cite{arana2016series,9646263}}\\
 \hline
 F$\&$R Actuator mass & 12\,kg\\
 F$\&$R Gear Ratio & 66 \\
 F/R Low speed shaft (LSS) & 166\,/\,165\,$N\!\!\cdot\!m$ \\
 continuous torque & \\
 F/R LSS peak torque & 273\,/\,273\,$N\!\!\cdot\!m$ \\
\hline
\hline
\\
\end{tabular*}}
\end{table}

\begin{table}[htb!]
\centering
\caption{PID tuning parameters in `PALS-PID' and `FF-PID-non' control schemes}\label{tab1-2}
\scalebox{1}{\begin{tabular*}{1\columnwidth}{@{\extracolsep{\fill}}l @{\extracolsep{\fill}} c@{\extracolsep{\fill}}
c@{\extracolsep{\fill}}c @{\extracolsep{\fill}} c@{\extracolsep{\fill}} c@{\extracolsep{\fill}}c}
\hline
\hline
Controller & Aim & Axle & P & I & D\\
 \hline
`PALS-PID' & pitch & F$\&$R & 1000 & 20000 & 4\\
 & roll & F$\&$R & 500 & 5000 & 4\\
`FF-PID-non' & pitch & F$\&$R & 100 & 2500 & 2\\
& roll & F$\&$R & 50 & 1500 & 2\\
\hline
\hline
\end{tabular*}}
\end{table}

\bibliographystyle{ieeetr}
\bibliography{reference}

\end{document}